# Continuous Authentication Using Mouse Clickstream Data Analysis

Sultan Almalki[1], Prosenjit Chatterjee[2] and Kaushik Roy[2]

[1, 2, 3]Department of Computer Science, North Carolina A&T State University, USA
[1]ssalmalki@aggies.ncat.edu, [2]pchatterjee@aggies.ncat.edu, [2]kroy@ncat.edu

**Abstract.** Biometrics is used to authenticate an individual based on physiological or behavioral traits. Mouse dynamics is an example of a behavioral biometric that can be used to perform continuous authentication as protection against security breaches. Recent research on mouse dynamics has shown promising results in identifying users; however, it has not yet reached an acceptable level of accuracy. In this paper, an empirical evaluation of different classification techniques is conducted on a mouse dynamics dataset, the Balabit Mouse Challenge dataset. User identification is carried out using three mouse actions: mouse move, point and click, and drag and drop. Verification and authentication methods are conducted using three machine-learning classifiers: the Decision Tree classifier, the K-Nearest Neighbors classifier, and the Random Forest classifier. The results show that the three classifiers can distinguish between a genuine user and an impostor with a relatively high degree of accuracy. In the verification mode, all the classifiers achieve a perfect accuracy of 100%. In authentication mode, all three classifiers achieved the highest accuracy (ACC) and Area Under Curve (AUC) from scenario B using the point and click action data: (Decision Tree ACC:87.6%, AUC:90.3%), (K-Nearest Neighbors ACC:99.3%, AUC:99.9%), and (Random Forest ACC:89.9%, AUC:92.5%).

## 1 Introduction

User authentication is a method that is used to determine whether a user is genuine ("allowed to access the system") or an impostor ("prohibited from access to the system") [1]. User authentication has three types of classes: knowledge based, object or token based, and biometric based. Knowledge-based user authentication is characterized by confidentiality; it is something that only the user would know. Object-based user authentication is characterized by control; it is something that the user has. Biometric-based user authentication relies on the user's physiological or behavioral characteristics; it is something the user is. While the weaknesses of knowledge-based and object-based approaches are that the user may lose or forget passwords and tokens, the advantage of a biometric-based approach is that it can uniquely identify an individual by using the individual's biological characteristics.

Although using biometrics makes the authentication stronger and determines a user's identity uniquely, verification based on physiological biometrics such as iris, face, or fingerprint offers mainly a one-time static authentication [2,3]. To avoid this drawback, behavioral biometrics such as mouse clickstream data can be used to continuously authenticate a user by monitoring the user's behavior [4]. In this work, an empirical evaluation of three classifiers is conducted on the Balabit dataset, which contains data for 10 users with a set of 39 behavioral features per user [5].

The rest of the paper is organized in four sections. Section 2 summarizes some previous research in this area. Section 3 describes the Balabit dataset and the feature extraction method. Section 4 describes the model and the experiments, followed by a discussion of the test results. Section 5 has concluding remarks and suggestions for future work.

## 2      Related Work

User behavioral analysis has been a focus of research for more than a decade. This section briefly presents some of the research on mouse-based authentication.

Antal et al. (2018) [5] applied a Random Forest (RF) classifier for each user using mouse movements for verifying impostor detection. They used the Balabit dataset, which includes 10 users. Each user has many sections and mouse actions. They segmented each session's data into three types of mouse actions: Mouse Movement (MM), Point Click (PC), and Drag and Drop (DD). The researchers extracted 39 features and obtained results of 80.17% average accuracy (ACC) and 0.87 average Area Under Curve (AUC). The highest accuracies achieved for users (7 and 9) were 93% and 0.97 AUC. The lowest accuracy achieved for a user (8) was 72% and 0.80 AUC.

Nakkabi et al. (2010) [6] proposed a user authentication scheme based on mouse dynamics. They collected mouse behavior data from 48 users and applied a fuzzy classification that relied on a learning algorithm for multivariate data analysis. They conducted an evaluation and achieved a False Acceptance Rate (FAR) of 0% and a False Rejection Rate (FRR) of 0.36%. Their experiments required more than 2000 mouse events in order to classify a user as legitimate.

Feher et al. (2012) [7] introduced a framework for user verification using mouse activities. The framework was divided into three parts: acquisition, learning, and verification. The first step is to capture user actions from the users' mouse activities. Then, classify each event type and store them in a database. The third phase is to send each event to the favorite classifier based on action type. The classifier has two layers: a prediction layer and a decision layer. The researchers conducted tests of multi-class classifier using a RF classifier. They collected the data from 25 volunteers. They obtained an Equal Error Rate (EER) of 1.01 % based on 30 actions.

Zheng et al. (2011) [8] proposed an approach of mouse movements for user verification. They collected data from 30 users with different ages, educational backgrounds, and occupations. The system utilized the angle-based metrics and Support Vector Machine (SVM). The results showed an EER of 1.3% that relied on 20 mouse clicks.

Another biometric authentication approach based on mouse dynamics was introduced by Shen et al. (2012) [9]. They collected user behavioral data under a controlled environment using the software tool they developed. The software collected the events of "mouse move" or "mouse click" for about thirty minutes in each session. The dataset obtained had 15 sessions for each of 28 subjects. Based on a mining method, the researchers focused on using frequent and fixed actions as behavioral patterns for extracting user characteristics through pattern growth. They used an SVM and achieved an FAR of 0.37% and an FRR of 1.12%.

Schulz (2006) [10] collected a dataset from 72 volunteers using a software tool on their personal machines. The software tool presented a continuous authentication system using mouse events; it segmented a user's events into length of a movement, curvature, inflection, and curve straightness features, and then computed a user's behavioral signature using histograms based on curve characteristics. For the verification stage, the researcher used Euclidean distance for classification and computed the distance between a user's login and the mouse activities. An EER of 24.3% from a group of 60 mouse curves is obtained. In contrast, by using groups of 3600 mouse curves, the performance increased to an EER of 11.2%.

Bours et al.in (2009) [11] proposed a login system based on mouse dynamics. They collected data from 28 participants of different age groups. They used a technique called "follow the maze" in which the participants performed a task by following the tracks on their own computer. This task was performed five times per session in order to acquire sufficient data on mouse movements. The maze contained 18 tracks, divided into 9 horizontal and 9 vertical tracks. They measured the various distances using Euclidean distance, Manhattan distance, and edit distance algorithms. The results that they obtained were an EER of 26.8% in the case of the horizontal direction and an EER=27.0% in the case of the vertical direction.

## 3     Description of Mouse Raw Data

This research used the Balabit Mouse Challenge dataset, obtained at the Budapest office of the Balabit company. The dataset contains raw data obtained from 10 users using remote desktop clients connected to remote servers. It has many sessions with characteristics of how a person uses a mouse. Each session includes a set of rows, where each row recorded a user action as (rtime, ctime, button, state, x, y): "rtime" is the elapsed time recorded since the start of the session using the network monitoring device, "ctime" is the elapsed time through the client computer, "button" is a mouse button, "state" is information about the button, and "x" and "y" are the Cartesian coordinates of the mouse location [5].

### 3.1 Extraction of Features

A mouse action is a set of sequential user actions that represent a movement of the mouse between two points. This study uses the user features extracted from the Balabit Mouse Challenge dataset [5]. This dataset divides the raw data into three types of actions: MM, PC, and DD. MM describes a movement between two screen positions; PC is a Point Click or Mouse click; DD is a drag-and-drop event. The dataset presents 39 features extracted from an individual's mouse actions. Table 1 shows the 39 features and their descriptions.

**Table 1.** Extraction of Features [Margit et al., 2018]

| Ranking | Name | Description |
|---|---|---|
| 1 | Type_of_action | Mouse Movement, Point Click, or Drag and Drop |
| 2 | Travelled_distance_in pixels | The frequency of actions within different distance ranges |
| 3 | Elapsed_time | Elapsed time from the start of the session recorded by the network monitoring device |
| 4 | Direction of movement | Direction of end to end line |
| 5 | Straightness | The ratio between two endpoints of action and the length of the trajectory |
| 6 | Num_points | Number of mouse events contained in an action |
| 7 | Sum_of_angles | How many angles in each action |
| 8 | Mean_curv | Average of angle change and the travelled distance |
| 9 | Sd_curv | Standard deviation between angle change and the travelled distance |
| 10 | Max_curv | Maximal values between angle change and the travelled distance |
| 11 | Min_curv | Minimal values between angle change and the travelled distance |
| 12 | Mean_omega | Average of angular velocity |
| 13 | Sd_omega | standard deviation of angular velocity |
| 14 | Max_ omega | maximal values of angular velocity |
| 15 | Min_omega | minimal values of angular velocity |
| 16 | Largest_deviation | largest distance between the points of the trajectory |
| 17 | Dist_end_to_end_line | Distance between two endpoints |
| 18 | Num_critical_points | Number of critical points |
| 19 | Mean_vx | Average of horizontal velocity |
| 20 | Sd_vx | Standard deviation of horizontal velocity |
| 21 | Max_vx | Maximal values of horizontal velocity |
| 22 | Min_vx | Minimal values of horizontal velocity |
| 23 | Mean_vy | Average of vertical velocity |
| 24 | Sd_vy | Standard deviation of vertical velocity |
| 25 | Max_vy | Maximal values of vertical velocity |
| 26 | Min_vy | Minimal values of vertical velocity |
| 27 | Mean_v | Average of velocity |

| Ranking | Name | Description |
|---------|------|-------------|
| 28 | Sd_v | Standard deviation of velocity |
| 29 | Max_v | Maximal values of velocity |
| 30 | Min_v | Minimal values of velocity |
| 31 | Mean_a | Average of acceleration |
| 32 | Sd_a | Standard deviation of acceleration |
| 33 | Max_a | Maximal values of acceleration |
| 34 | Min_a | Minimal values of acceleration |
| 35 | Mean_ jerk | Average of jerk |
| 36 | Sd_jerk | Standard deviation of jerk |
| 37 | Max_jerk | Maximal values of jerk |
| 38 | Min_jerk | Minimal values of jerk |
| 39 | A_beg_time | Acceleration of time at the beginning |

## 4    Mouse Dynamics Model and Experimental Results

In this research, supervised machine-learning techniques were utilized to monitor the behavior of users in order to distinguish legal users from illegal users. Three machine-learning algorithms were evaluated: Decision Tree Learning (DT), k-Nearest Neighbors (k-NN), and Random Forest (RF). The Scikit-learn software tools were used for the analysis of mouse clickstream data [12]. A significant step in the classification was to prepare the training data in CSV format, so that it could be interpreted by the classifiers. In the model, if a user's mouse dynamics are the same as the characteristics stored in the system's database, then the system lets the user continue working on the device; otherwise, the system must log out the user ( Fig. 1). Specifically, the following steps describe how the model works:

- Data Collection Phase: Raw data of the users are collected.
- Features Extraction Phase: Meaningful features, such MM, PC, and DD, were extracted using the method reported in Antal et al. [5].
- Data Preparation Phase: For the training phase, all the users' data was aggregated and put in random order. The training dataset was then split into two parts: the first part (70% of the data) was used for training, and the second part (30% of the data) was used for testing the model's performance. For every experiment, the balance of training sets and evaluation sets remained the same in order to avoid classifier bias.
- Select a Classifier Phase: DT, RF, and KNN were utilized to show the ability of the proposed model to determine whether a user was genuine or an impostor from a user's mouse clickstream data.
- Training Data Phase: The training process began by reading the characteristics of all the users from the training dataset and then loading them into the three classifiers to train the model. This step was a significant step, since the training data contained the user behavior itself and a class label.

6- Testing Data Phase: After completion of the training step, the model was tested on the new data that was never used for training, to categorize whether the user as a genuine user or an impostor.

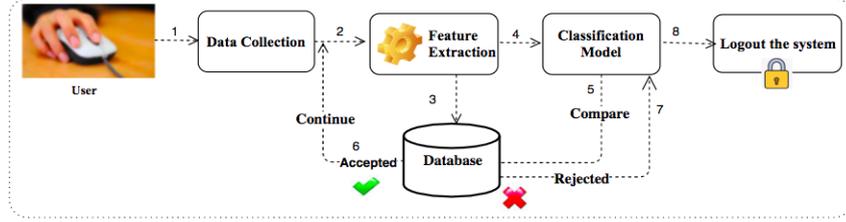

**Fig. 1.** User Behavioral Biometrics Model

The experiment was conducted in two stages: (i) a verification stage, and (ii) an authentication stage. The evaluations were measured using classifier accuracy (ACC) and area under curve (AUC). Evaluation results are reported in terms of false acceptance rate(FAR) , false rejection rate and equal error rate (FAR) . FAR is a measure of the chance that a user who should be rejected is accepted by the system. FRR is a measure of the chance that a user who should be accepted is rejected by the system. ERR is a threshold value between the false acceptance rate and the false rejection rate. Another important evaluation to examine the classifiers is to plot the receiver operating characteristic (ROC). The ROC curve plots the True Positive Rate(TPR) against the False Positive Rate (FPR). The following expressions are used for performance evaluations purposes [13,14]:

$$ACC = \frac{TP + TN}{TP + TN + FP + FN} \qquad (1)$$

$$TPR = \frac{TP}{TP + FN} \qquad (2)$$

$$TNR = \frac{TN}{TN + FP} \qquad (3)$$

$$FPR = \frac{FP}{FP + TN} \qquad (4)$$

$$FNR = \frac{FN}{FN + TP} \qquad (5)$$

$$FAR = \frac{\text{Number of accepted impostors}}{\text{Tatal number of impostors}} \qquad (6)$$

$$FRR = \frac{\text{Number of rejected genuines}}{\text{Tatal number of genuines}} \qquad (7)$$

$$EER = \frac{FAR + FRR}{2} \qquad (8)$$

Where, TP: True Positive, TN: True Negative, FP: False Positive and FN: False Negative.

### 4.1 Verification Stage

In this stage, all three classifiers were first trained using the data that only contained the genuine user's actions (positive). Each user has many sessions; all users' sessions data were placed in one Excel file. Then, the experiment was conducted by doing training and testing for each user using the DT, K-NN, and RF classifiers. The goal of the verification stage was to verify whether the mouse data was related to a given user. After testing all the users using three classifiers, a perfect score of 100% verification rate was achieved.

### 4.2 Authentication Stage

In this stage, each user is in one of two classes: genuine (positive) and impostor (negative). The impostor actions were selected from the other users. The classifiers are responsible for determining the probability that the user belongs to the genuine class or imposter class . Therefore, all classifiers were tested based on these two scenarios:

   A.  A single user's data with all actions (MM, PC, DD)
   B.  All the users' data with a single action (MM, PC, DD)

**Scenario A: A Single User's Data with All Actions.** In scenario (A), an experiment was conducted for a single user (7, 9, 12, 15, 16, 20, 21, 23, 29, and 35) with all actions (MM, PC, and DD), using the three classifiers. The DT, K-NN, and RF classifiers achieved average accuracies of 91.9%, 94.4%, and 79.7%, respectively. The highest average accuracies were achieved for user (9): (ACC: 91.8%), DT 96.2%, KNN 99.2%, and RF 80.1%. The lowest average accuracies were achieved for user (12): (ACC: 85.6%), DT 90.1%, KNN 91.5%, and RF 75.2%. Table 2 reports the results in detail for each user. The AUC value is computed based on the FPR and the TPR.

Table 2. Scenario A: Single user, all actions (MM, PC, DD)

| User | Decision Tree | | K-Nearest Neighbors | | Random Forest | |
|---|---|---|---|---|---|---|
|  | ACC% | AUC | ACC% | AUC | ACC% | AUC |
| 35 | 84.9 | 92.1 | 96.6 | 99.4 | 88.3 | 91.2 |
| 7  | 92.4 | 93.8 | 88.7 | 92.2 | 85.8 | 88.1 |
| 9  | 96.2 | 97.1 | 99.2 | 99.1 | 80.1 | 81.0 |
| 12 | 90.1 | 97.5 | 91.5 | 99.2 | 75.2 | 79.7 |
| 15 | 92.6 | 98.1 | 99.7 | 99.3 | 80.5 | 82.5 |
| 16 | 88.6 | 91.0 | 97.3 | 99.4 | 84.9 | 86.7 |
| 20 | 93.8 | 97.2 | 90.1 | 99.0 | 75.6 | 80.5 |
| 21 | 95.6 | 97.9 | 92.4 | 99.3 | 72.8 | 77.3 |
| 23 | 91.1 | 96.4 | 95.2 | 99.3 | 82.2 | 84.9 |
| 29 | 94.5 | 96.5 | 93.5 | 99.8 | 71.7 | 74.4 |
| Avg | 91.9 | 95.7 | 94.4 | 98.6 | 79.7 | 82.6 |

**Scenario B**: **All Users' Data with a Single Action.** In scenario (B), the dataset was initially separated into three groups of mouse actions: MM, PC, and DD. Each group contained all users (7, 9, 12, 15, 16, 20, 21, 23, 29, and 35). Training and testing of the three classifiers were then conducted on each group based on the single action. The results are reported in Table 3 (MM), Table 4 (PC), and Table 5 (DD). The highest accuracies were achieved with the PC action compared to MM and DD, as shown in Table 4 (PC): (DT: ACC:87.6%, AUC:90.3%), (KNN: ACC:99.3%, AUC:99.9%), and (RF: ACC:89.9%, AUC:92.5%).

**Table 3.** Scenario B: All users, single action (MM action)

| User | Decision Tree | | K-Nearest Neighbors | | Random Forest | |
|---|---|---|---|---|---|---|
|  | ACC% | AUC | ACC% | AUC | ACC% | AUC |
| 35 | 92.9 | 95.8 | 99.5 | 100 | 97.3 | 99.0 |
| 7  | 95.4 | 98.1 | 99.7 | 100 | 98.8 | 99.8 |
| 9  | 83.2 | 86.7 | 99.2 | 99.9 | 85.1 | 87.6 |
| 12 | 81.1 | 84.0 | 99.5 | 99.6 | 86.2 | 89.9 |
| 15 | 80.6 | 83.0 | 99.7 | 99.9 | 88.5 | 91.9 |
| 16 | 93.6 | 96.3 | 99.3 | 99.8 | 93.9 | 95.2 |
| 20 | 80.8 | 84.4 | 99.1 | 100 | 87.6 | 90.7 |
| 21 | 78.6 | 80.6 | 99.4 | 99.6 | 80.8 | 84.5 |
| 23 | 75.7 | 78.1 | 99.2 | 99.7 | 85.2 | 89.6 |
| 29 | 79.5 | 81.2 | 99.5 | 99.4 | 82.7 | 85.3 |
| Avg | 84.1 | 86.8 | 99.4 | 99.8 | 88.6 | 91.3 |

**Table 4.** Scenario B: All users, single action (PC action)

| User | Decision Tree | | K-Nearest Neighbors | | Random Forest | |
|---|---|---|---|---|---|---|
|  | ACC% | AUC | ACC% | AUC | ACC% | AUC |
| 35 | 93.9 | 95.7 | 98.6 | 99.9 | 91.3 | 94.4 |
| 7  | 95.4 | 97.6 | 99.7 | 100 | 98.8 | 99.7 |
| 9  | 85.2 | 88.7 | 99.2 | 100 | 89.1 | 92.4 |
| 12 | 90.1 | 93.4 | 99.5 | 99.9 | 86.2 | 89.9 |
| 15 | 84.6 | 86.5 | 99.7 | 99.9 | 88.5 | 91.0 |
| 16 | 91.6 | 94.8 | 99.3 | 100 | 95.9 | 97.1 |
| 20 | 86.8 | 89.1 | 99.1 | 99.9 | 88.6 | 91.4 |
| 21 | 82.6 | 85.0 | 99.9 | 99.9 | 89.1 | 91.0 |
| 23 | 83.1 | 87.8 | 99.2 | 99.8 | 89.2 | 92.3 |
| 29 | 82.5 | 84.7 | 98.9 | 99.8 | 82.7 | 85.5 |
| Avg | 87.6 | 90.3 | 99.3 | 99.9 | 89.9 | 92.5 |

Table 5. Scenario B: All users, single action (DD action)

| User | Decision Tree | | K-Nearest Neighbors | | Random Forest | |
|---|---|---|---|---|---|---|
| | ACC% | AUC | ACC% | AUC | ACC% | AUC |
| 35 | 92.3 | 94.5 | 98.6 | 99.4 | 98.3 | 99.0 |
| 7 | 93.9 | 95.5 | 95.7 | 97.9 | 95.8 | 97.8 |
| 9 | 82.5 | 86.9 | 98.2 | 99.7 | 87.1 | 91.8 |
| 12 | 85.3 | 89.3 | 98.5 | 99.5 | 89.2 | 93.5 |
| 15 | 88.1 | 90.5 | 99.7 | 100 | 90.5 | 93.1 |
| 16 | 87.6 | 89.6 | 98.3 | 99.6 | 91.9 | 94.4 |
| 20 | 85.8 | 88.2 | 98.1 | 99.5 | 89.6 | 92.1 |
| 21 | 85.6 | 89.2 | 96.4 | 98.2 | 79.8 | 82.8 |
| 23 | 85.2 | 87.8 | 98.2 | 99.5 | 93.2 | 96.0 |
| 29 | 82.8 | 85.0 | 98.5 | 99.6 | 80.7 | 84.4 |
| Avg | 86.9 | 89.7 | 98.0 | 99.3 | 89.6 | 92.5 |

In the following sections, evaluation results are provided for scenarios A and B in terms of FAR, FRR, and EER. ROC curves are also given.

**Scenario A: Single user, all actions (MM, PC, and DD) – additional information.** This scenario is a single user with all actions (MM, PC, DD). Both positive and negative actions were used to evaluate the classifiers. The averages of FARs for all users are (DT:0.007, KNN:0.003, RF:0.052). The averages of FRRs for all users are (DT:0.077, KNN:0.029, RF:0.473). The averages of EERs for all users are (DT:0.070, KNN:0.012,RF:0.247). Table 6 shows the results for all users. ROC curves are given in Fig. 2, **Fig. 3**, and Fig. 4.

Table 6. FAR, FRR, and EER - Scenario A - single user, all actions (MM, PC, DD)

| User | Decision Tree | | | K-Nearest Neighbors | | | Random Forest | | |
|---|---|---|---|---|---|---|---|---|---|
| | FAR | FRR | EER | FAR | FRR | EER | FAR | FRR | EER |
| 35 | 0.015 | 0.146 | 0.129 | 0.003 | 0.019 | 0.008 | 0.037 | 0.224 | 0.140 |
| 7 | 0.007 | 0.121 | 0.109 | 0.002 | 0.027 | 0.012 | 0.017 | 0.343 | 0.185 |
| 9 | 0.006 | 0.042 | 0.041 | 0.004 | 0.028 | 0.011 | 0.092 | 0.426 | 0.263 |
| 12 | 0.005 | 0.040 | 0.039 | 0.004 | 0.030 | 0.014 | 0.101 | 0.462 | 0.259 |
| 15 | 0.006 | 0.036 | 0.035 | 0.005 | 0.030 | 0.012 | 0.092 | 0.403 | 0.403 |
| 16 | 0.011 | 0.181 | 0.155 | 0.002 | 0.020 | 0.005 | 0.009 | 0.540 | 0.198 |
| 20 | 0.005 | 0.056 | 0.053 | 0.002 | 0.039 | 0.017 | 0.038 | 0.576 | 0.255 |
| 21 | 0.003 | 0.041 | 0.039 | 0.003 | 0.029 | 0.009 | 0.038 | 0.581 | 0.267 |
| 23 | 0.006 | 0.056 | 0.053 | 0.003 | 0.025 | 0.011 | 0.047 | 0.433 | 0.202 |
| 29 | 0.005 | 0.056 | 0.053 | 0.001 | 0.049 | 0.022 | 0.053 | 0.744 | 0.305 |
| Avg | 0.007 | 0.077 | 0.070 | 0.003 | 0.029 | 0.012 | 0.052 | 0.473 | 0.247 |

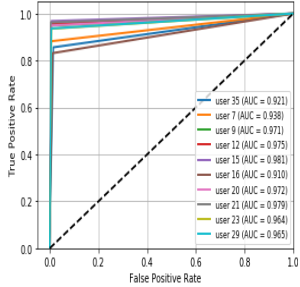 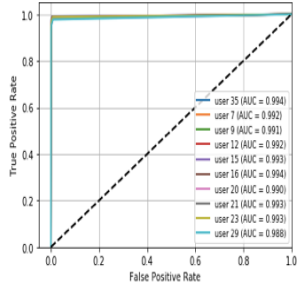 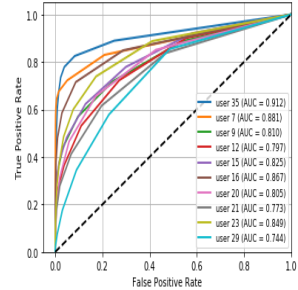

**Fig. 2.** ROC curve for DT, single user, all actions  **Fig. 3.** ROC curve for KNN, single user, all actions  **Fig. 4.** ROC curve for RF, single user, all actions

**Scenario B: All users, single action – additional information.** In this scenario, the experiments were conducted on all user data using one type of mouse action (MM, PC, DD). We trained and tested all the users' data in both positive and negative actions and evaluated the three classifiers. We should note that all classifiers were evaluated for each action separately. In the following sections, we report evaluation results for all classifiers in each action:

For the MM action, the averages of FARs for all users are (DT FAR:0.053,KNN FAR:0.006,RF FAR:0.045).The averages of FRRs for all users are (DT FRR:0.455, KNN FRR:0.075, RF FRR:0.416). The averages of EERs for all users are (DT EER:0.216, KNN ERR:0.011, RF ERR:0.173). The results for all users are shown in Table 7. ROC curves are shown in Fig.5, Fig.6 and Fig. **7**.

**Table 7.** FAR, FRR, and EER - Scenario B - all users (MM action)

| User | Decision Tree | | | K-Nearest Neighbors | | | Random Forest | | |
|---|---|---|---|---|---|---|---|---|---|
| | FAR | FRR | EER | FAR | FRR | EER | FAR | FRR | EER |
| 35 | 0.017 | 0.280 | 0.091 | 0.006 | 0.001 | 0.005 | 0.030 | 0.049 | 0.041 |
| 7 | 0.017 | 0.242 | 0.245 | 0.006 | 0.128 | 0.007 | 0.004 | 0.099 | 0.036 |
| 9 | 0.059 | 0.380 | 0.019 | 0.013 | 0.015 | 0.014 | 0.088 | 0.372 | 0.213 |
| 12 | 0.001 | 0.663 | 0.276 | 0.007 | 0.040 | 0.016 | 0.046 | 0.489 | 0.226 |
| 15 | 0.013 | 0.573 | 0.284 | 0.011 | 0.001 | 0.006 | 0.098 | 0.303 | 0.197 |
| 16 | 0.022 | 0.252 | 0.103 | 0.001 | 0.343 | 0.015 | 0.012 | 0.475 | 0.105 |
| 20 | 0.053 | 0.514 | 0.266 | 0.005 | 0.006 | 0.008 | 0.090 | 0.357 | 0.198 |
| 21 | 0.001 | 0.715 | 0.303 | 0.006 | 0.003 | 0.006 | 0.017 | 0.796 | 0.262 |
| 23 | 0.004 | 0.892 | 0.312 | 0.006 | 0.089 | 0.025 | 0.039 | 0.475 | 0.187 |
| 29 | 0.345 | 0.047 | 0.265 | 0.001 | 0.124 | 0.015 | 0.033 | 0.736 | 0.265 |
| Avg | 0.053 | 0.455 | 0.216 | 0.006 | 0.075 | 0.011 | 0.045 | 0.416 | 0.173 |

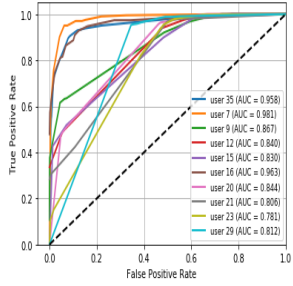
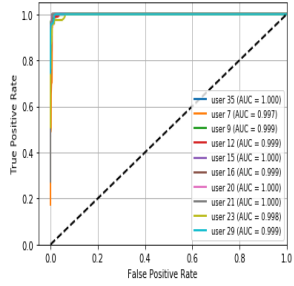
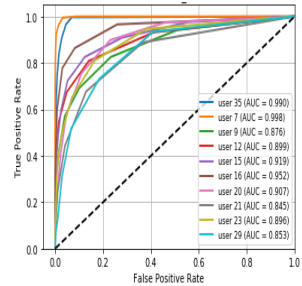

**Fig. 5.** ROC curve for DT, all users, MM action

**Fig. 6.** ROC curve for KNN, all users, MM action

**Fig. 7.** ROC curve for RF, all users, MM action

For the PC action, the averages of FARs for all users are (DT FAR:0.049, KNN FAR:0.846, RF FAR:0.040). The averages of FRRs for all users are (DT FRR: 0.446, KNN FRR:0.005, RF FAR:0.368). The averages of EERs for all users are (DT EER: 0.186, KNN EER:0.847, RF EER:0.152). The detailed results are shown Table 8. ROC curves are shown in Fig. **8**, Fig.9, and Fig. **10**.

**Table 8.** FAR, FRR, and EER - Scenario B - all users (PC action)

| User | Decision Tree | | | K-Nearest Neighbors | | | Random Forest | | |
|---|---|---|---|---|---|---|---|---|---|
| | FAR | FRR | EER | FAR | FRR | EER | FAR | FRR | EER |
| 35 | 0.039 | 0.198 | 0.098 | 0.002 | 0.001 | 0.001 | 0.029 | 0.046 | 0.042 |
| 7 | 0.010 | 0.212 | 0.086 | 8.444 | 0.001 | 8.444 | 0.005 | 0.099 | 0.022 |
| 9 | 0.009 | 0.527 | 0.211 | 0.001 | 0.001 | 0.002 | 0.078 | 0.339 | 0.203 |
| 12 | 0.044 | 0.230 | 0.150 | 0.001 | 0.004 | 0.008 | 0.034 | 0.382 | 0.175 |
| 15 | 0.314 | 0.153 | 0.249 | 0.003 | 0.005 | 0.001 | 0.083 | 0.221 | 0.156 |
| 16 | 0.003 | 0.491 | 0.120 | 0.002 | 0.007 | 0.003 | 0.006 | 0.490 | 0.116 |
| 20 | 0.057 | 0.423 | 0.215 | 0.001 | 0.008 | 0.001 | 0.075 | 0.337 | 0.171 |
| 21 | 0.001 | 0.810 | 0.251 | 0.001 | 0.003 | 0.003 | 0.019 | 0.692 | 0.231 |
| 23 | 0.019 | 0.654 | 0.224 | 0.002 | 0.008 | 0.003 | 0.037 | 0.422 | 0.180 |
| 29 | 0.001 | 0.770 | 0.261 | 0.001 | 0.020 | 0.002 | 0.035 | 0.659 | 0.228 |
| Avg | 0.049 | 0.446 | 0.186 | 0.846 | 0.005 | 0.847 | 0.040 | 0.368 | 0.152 |

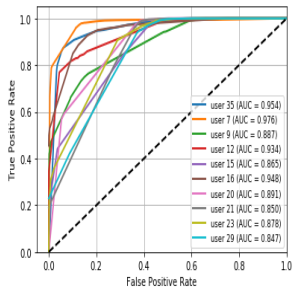
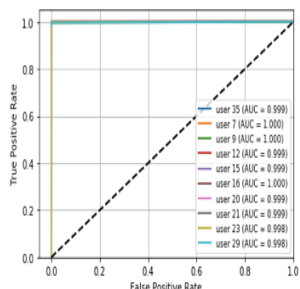
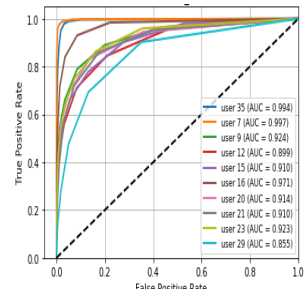

**Fig. 8.** ROC curve for DT, all users, PC action

**Fig. 9.** ROC curve for KNN, all users, PC action

**Fig. 10.** ROC curve for RF, all users, PC action

For the DD action, the averages of FARs for all users are (DT FAR:0.053, KNN FAR:0.021, RF FAR:0.033). The averages of FRRs for all users are (DT FRR:0.517, KNN FRR:0.246, RF FRR:0.363). The averages of EERs for all users are (DT EER:0.186, KNN EER:0.021, RF EER:0.138). The detailed results are shown in Table 7. ROC curves are shown in Fig.**11**, Fig. **12**, and Fig. **13**.

**Table 9.** FAR, FRR, and EER - Scenario B - all users (DD action)

| User | Decision Tree | | | K-Nearest Neighbors | | | Random Forest | | |
|---|---|---|---|---|---|---|---|---|---|
|  | FAR | FRR | EER | FAR | FRR | EER | FAR | FRR | EER |
| 35 | 0.039 | 0.303 | 0.098 | 0.003 | 0.019 | 0.026 | 0.029 | 0.091 | 0.049 |
| 7 | 0.025 | 0.212 | 0.086 | 0.002 | 1.000 | 0.012 | 0.004 | 0.231 | 0.040 |
| 9 | 0.212 | 0.258 | 0.220 | 0.065 | 0.001 | 0.011 | 0.065 | 0.262 | 0.155 |
| 12 | 0.001 | 0.598 | 0.209 | 0.034 | 0.002 | 0.001 | 0.035 | 0.314 | 0.126 |
| 15 | 0.170 | 0.220 | 0.199 | 0.005 | 0.030 | 0.012 | 0.090 | 0.177 | 0.138 |
| 16 | 0.029 | 0.350 | 0.156 | 0.042 | 0.033 | 0.011 | 0.005 | 0.39 | 0.107 |
| 20 | 0.006 | 0.862 | 0.215 | 0.011 | 0.309 | 0.017 | 0.039 | 0.425 | 0.189 |
| 21 | 0.001 | 0.943 | 0.199 | 0.003 | 0.775 | 0.069 | 0.012 | 0.820 | 0.236 |
| 23 | 0.007 | 0.654 | 0.224 | 0.043 | 0.065 | 0.043 | 0.040 | 0.192 | 0.092 |
| 29 | 0.040 | 0.770 | 0.261 | 0.007 | 0.232 | 0.016 | 0.012 | 0.732 | 0.254 |
| Avg | 0.053 | 0.517 | 0.186 | 0.021 | 0.246 | 0.021 | 0.033 | 0.363 | 0.138 |

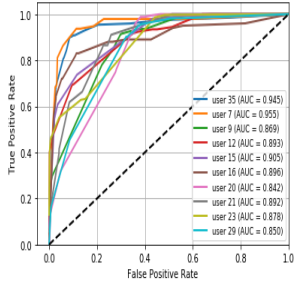

**Fig. 11.** ROC curve for DT, all users, DD action

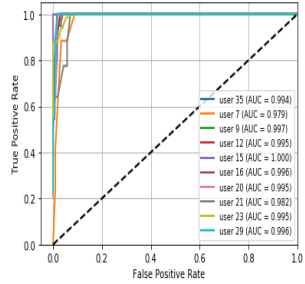

**Fig. 12.** ROC curve for KNN, all users, DD action

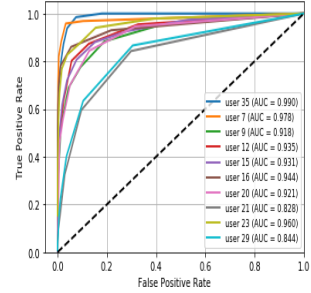

**Fig. 13.** ROC curve for RF, all users, DD action

## 5    Conclusion

This paper provides a continuous user authentication model based on mouse click-stream data analysis. Each of three machine-learning classifiers used 39 features of mouse actions MM, PC, and DD. The classifiers were able to determine a genuine user from an impostor with reasonable accuracies and AUC.

In the verification phase, the model was able to recognize the user with an accuracy of 100%. In the authentication phase, data containing genuine and impostor actions were examined using two scenarios: (A) a single user with all actions, and (B) a single action with all users. The best results were obtained from scenario B using the PC action: (DT - ACC: 87.6%, AUC: 90.3%), (KNN - ACC: 99.3%, AUC: 99.9%), and (RF - ACC: 89.9%, AUC: 92.5%). In the future, a deep learning model will be constructed using the MM, PC, and DD actions, and its performance will be compared with the traditional classifiers.